\documentclass[aps,pra,twocolumn,superscriptaddress,nofootinbib]{revtex4}

\usepackage{graphicx}
\usepackage[normal]{subfigure}
\usepackage{caption}
\usepackage{latexsym}
\usepackage{amsmath}
\usepackage{amssymb}
\usepackage{amsfonts}
\usepackage{color}
\usepackage{pgf, tikz}
\usepackage{dsfont}
\usepackage{caption}
\usepackage{amsthm}
\usepackage{array}
\usepackage{tabulary}
\usepackage{mathtools}

\usepackage[utf8]{inputenc}
\usepackage[english]{babel}

\begin{document}

\newcommand{\ket}[1]{\vert #1 \rangle}
\newcommand{\bra} [1] {\langle #1 \vert}
\newcommand{\braket}[2]{\langle #1 | #2 \rangle}
\newcommand{\proj}[1]{\ket{#1}\bra{#1}}
\newcommand{\mean}[1]{\langle #1 \rangle}
\newcommand{\opnorm}[1]{|\!|\!|#1|\!|\!|_2}
\newtheorem{theo}{Theorem}
\newtheorem{lem}{Lemma}
\newtheorem{defin}{Definition}
\newtheorem{corollary}{Corollary}
 \newtheorem{conj}{Conjecture}
 \newtheorem*{prop}{Properties}
\newcommand{\mmmean}[1]{\langle\hspace{-3pt} \langle\hspace{-3pt} \langle #1 \rangle\hspace{-3pt} \rangle\hspace{-3pt} \rangle}
\newcommand{\mmean}[1]{\langle\hspace{-3pt} \langle #1 \rangle\hspace{-3pt} \rangle}
 \newcommand{\kket}[1]{\vert\hspace{-2pt}\vert #1 \rangle \hspace{-3pt}\rangle}
\newcommand{\bbra} [1] {\langle \hspace{-3pt} \langle #1 \vert\hspace{-2pt}\vert}
\newcommand{\eq}[1]{\begin{equation}#1	\end{equation}}
\newcommand{\eqarray}[1]{\begin{eqnarray}#1	\end{eqnarray}}
\newcommand{\red}[1]{\textcolor{red}{#1}}
\newcommand{\x}{\hat{x}_1}
\newcommand{\xx}{\hat{x}_2}
\newcommand{\p}{\hat{p}_1}
\newcommand{\pp}{\hat{p}_2}
\newcommand{\tr}{\mathrm{Tr}}


\title{Multi-copy uncertainty observable inducing a symplectic-invariant uncertainty relation\\
in position and momentum phase space}

\author{Anaelle Hertz}
\affiliation{Centre for Quantum Information and Communication, \'Ecole polytechnique de Bruxelles, CP 165, Universit\'e libre de Bruxelles, 1050 Brussels, Belgium}

\author{Ognyan Oreshkov}
\affiliation{Centre for Quantum Information and Communication, \'Ecole polytechnique de Bruxelles, CP 165, Universit\'e libre de Bruxelles, 1050 Brussels, Belgium}

\author{Nicolas J. Cerf}
\affiliation{Centre for Quantum Information and Communication, \'Ecole polytechnique de Bruxelles, CP 165, Universit\'e libre de Bruxelles, 1050 Brussels, Belgium}

\begin{abstract}

We define an \textit{uncertainty observable}  acting on several replicas of a continuous-variable state, whose measurement induces phase-space uncertainty relations for a single copy of the state. By exploiting the Schwinger representation of angular momenta in terms of bosonic operators, this observable can be constructed so as to be invariant under symplectic transformations (rotation and squeezing in phase space). We first  design a \textit{two-copy} uncertainty observable, which is a discrete-spectrum operator vanishing with certainty if and only if it is applied on (two  replicas of) any pure Gaussian state centered at the origin. The non-negativity of its variance translates into the Schrödinger-Robertson uncertainty relation. We then extend our construction to a \textit{three-copy} uncertainty observable, which exhibits additional invariance under displacements (translations in phase space) so that it vanishes on every pure Gaussian state. The resulting invariance under  all Gaussian unitaries makes this observable a natural tool to capture the phase-space uncertainty -- or the deviation from pure Gaussianity -- of continuous-variable bosonic states. In particular, it suggests that the Shannon entropy  associated with the measurement of this observable provides a symplectic-invariant entropic measure of uncertainty  in position-momentum phase space.

\end{abstract}

\maketitle



The seminal uncertainty relation due to Heisenberg \cite{Heisenberg}  and more precisely formulated by Kennard \cite{kennard} states that 
\begin{equation}
\Delta x^2\, \Delta p^2\geq 1/4 \, , 
\label{Schrodinger-Kennard}
\end{equation}
where $\Delta x^2$ and $\Delta p^2$ are the position and momentum variances ($\hbar=1$). The set of states that saturate this uncertainty relation are all pure Gaussian states whose covariance vanishes, i.e., those that have no $x$-$p$ correlation (see \cite{hertz2}). Other pure Gaussian states are not minimum-uncertainty states according to the measure  implied by the left-hand side of Heisenberg relation (\ref{Schrodinger-Kennard}) as a consequence of the fact that the latter is not invariant under rotations in phase space. This problem was solved by Schrödinger \cite{Schrodinger} and Robertson \cite{Robertson}, who added an anticommutator term giving rise to the uncertainty relation 
\begin{equation}
\det\gamma\geq 1/4 \, , 
\label{Schrodinger-Robertson-UR}
\end{equation}
with $\gamma$ being the covariance matrix.
Since this determinant is invariant under symplectic transformations (rotation and squeezing) as well as displacements (translations) \cite{RMP}, and since it reduces to $\Delta x^2\, \Delta p^2$ for states with vanishing covariance, the Schrödinger-Robertson relation~(\ref{Schrodinger-Robertson-UR}) is saturated by all pure Gaussian states, which form the set of minimum-uncertainty states.

Variances, however, are not the only possible measure of uncertainty. In information theory, a much preferred quantity is the Shannon entropy. This measure can naturally also be applied to expressing uncertainty relations. Bialynicki-Birula and Mycielski \cite{birula} have indeed proven an entropic form of the uncertainty relation for continuous variables $x$ and $p$, namely 
\begin{equation}
h(x)+h(p)\geq\ln(\pi e)
\label{BBM}
\end{equation}
where $h(\cdot)$ stands for the Shannon differential entropy
\begin{equation}
h(x)=-\int p(x)\ln p(x) \, dx
\end{equation}
and $p(x)$ is the probability density function of $x$. In some sense, entropic uncertainty relations can be considered superior to variance-based uncertainty relations. For example, it is possible to derive Eq.~(\ref{Schrodinger-Kennard}) from Eq.~(\ref{BBM}), see \cite{hertz2}. The advent of quantum information theory and the special role played by entropies in this field also explains the renewed interest in entropic uncertainty relations over the last decade, see e.g. \cite{Bialynicki-Birula-Rudnicki,coles,Hertz-Cerf} for recent reviews. Note that  entropic uncertainty relations can be formulated for discrete variables as well, using the Shannon entropy
\begin{equation}
H(X)=-\sum_i p_i\ln p_i
\label{shannonentropy}
\end{equation}
where $p_i$ is the probability of measuring the outcome $x_i$. Here, the advantage over the Heisenberg or Schrödinger-Robertson relation is the possibility to obtain a state-independent uncertainty lower bound, see e.g. \cite{coles}.

A main drawback of the entropic uncertainty relation of Bialynicki-Birula and Mycielski is that its saturation is only reached for pure Gaussian states with zero covariance. This is because Eq.~(\ref{BBM}) is not invariant under rotations (or, more generally, symplectic transformations), exactly as Eq.~(\ref{Schrodinger-Kennard}). Recent progress has been made to define an entropic counterpart to the Schrödinger-Robertson relation \cite{hertz2}, but no symplectic-invariant uncertainty relation that is solely expressed in terms of entropies has been found as of today.
%
A possible, rather simple solution could be to consider the canonical pair of rotated variables $x_{\theta} = x \, \cos \theta  + p \, \sin \theta$ and $p_{\theta} = - x \, \sin \theta  + p \, \cos \theta$, where $\theta$ is a rotation angle. Then, one could take the average or even the minimum over $\theta$, giving respectively
\begin{equation}
{1\over 2 \pi} \int_0^{2\pi}  \left[ h(x_\theta)+h(p_\theta)\right]   \, d\theta  \ge \ln(\pi e)
\label{trial1}
\end{equation}
or
\begin{equation}
\min_\theta  \left[ h(x_\theta)+h(p_\theta)\right]   \, d\theta  \ge \ln(\pi e)
\label{trial2}
\end{equation}
This would apparently yield two variants of a symplectic-invariant uncertainty relation based on entropies (the latter being clearly stronger than the former). However, the quantities in the left-hand side of Eqs.~(\ref{trial1}) and (\ref{trial2}) do not appear easily tractable, so that the problem remains arguably open to define a useful entropic uncertainty relation that is invariant under symplectic transformations.




In this paper, we follow a path towards this goal consisting of enforcing the invariance of the measured observable instead of that of the uncertainty measure itself. We  develop a framework based on the Schwinger representation of angular momenta in terms of bosonic annihilation and creation operators. This enables us to define a multi-copy {\it uncertainty observable} with ingrained invariance under symplectic transformations in phase space (or under all Gaussian unitaries in continuous-variable state space). Then, measuring this observable allows us to express alternative uncertainty relations which logically have the appropriate invariance.

In Section 1.A, we define a {\it two-copy} uncertainty observable denoted as $\hat L_z$, which acts on two replicas of a bosonic state and is isomorphic to the $z$-component of an angular momentum. We present its physical representation  in Section 1.B and complete it with the other two components $\hat L_x$ and $\hat L_y$ in Section 1.C. The eigensystem of $\hat L_z$ is then analyzed in Section 1.D, where it is shown in particular that $\hat L_z$ takes on (half-) integer values from $-n/2$ to $n/2$ for a $n$-boson system. It is invariant under symplectic transformations (rotation and squeezing), and vanishes with probability one if and only if it is applied onto a Gaussian pure state that is centered at the origin in phase space. Remarkably, expressing the condition that this discrete-spectrum operator $\hat L_z$ has a non-negative variance translates into the Schrödinger-Robertson uncertainty relation based on the covariance matrix $\gamma$ for continuous variables $x$ and $p$. Then, in Section 1.E, we suggest that the Shannon entropy of $\hat L_z$ provides a hitherto unknown measure of uncertainty in phase space, which we compare to the Shannon differential entropy of the Wigner function in the special case of one-mode Gaussian states in Section 1.F.


Section 2 deals with the fact that $\hat L_z$ expresses an uncertainty only if applied to states centered at the origin. To overcome this limitation, we define in Section 2.A-2.B a {\it three-copy} uncertainty observable denoted as $\hat M$, which exhibits extra invariance under displacements (Weyl operators), hence admits all pure Gaussian states as minimum-uncertainty states. The resulting invariance under all Gaussian unitaries (rotation, squeezing, and displacement) makes this observable $\hat M$ a very natural measure of uncertainty in phase space -- or deviation from pure Gaussianity. Its spectrum is (one half) the spectrum of an angular momentum and, here too, the non-negativity of its variance coincides with the Schrödinger-Robertson uncertainty relation. The physical realization of the measurement of $\hat M$ is illustrated in Section 2.C. Then, in Section 2.D, we derive a symplectic-invariant entropic uncertainty relation based on the Shannon entropy of $\hat M$. It is shown that, for Gaussian states, the entropy of both multi-copy observables ($\hat L_z$ and $\hat M$) are equal. The case of non-Gaussian states is also briefly discussed. Finally, we give our conclusions in Section 3.

\section{Two-copy uncertainty observable}

\subsection{Definition of $\hat L_z$}
\label{sect-1A}

Let us gain intuition on how to define an uncertainty observable. In some vague sense, we are looking for an observable that could simultaneously access both $x$ and $p$ quadratures\footnote{From now on, we consider the $x$ and $p$ variables to be the canonically conjugate quadrature components of the electromagnetic field and adopt this quantum optics nomenclature. Our results, however, hold for any canonical pair of  variables that is analog to the position-momentum pair. }. To make it more precise, we consider a 2-copy observable which is acting on two identical copies of state $|\psi\rangle$.  Defining  $|\Psi\rangle \equiv |\psi\rangle_1 \otimes |\psi\rangle_2$ as the joint state of systems 1 and 2, we may simply consider the 2-copy observable $\hat O = \hat x_1 \otimes \hat p_2$. Its mean value gives 
\begin{equation}
\mmean{\hat O}_\Psi \equiv   \langle \Psi | \hat O | \Psi \rangle = \mean{\psi | \hat x | \psi} \, \mean{\psi | \hat p | \psi}
\end{equation}
where we will use, throughout this paper,  the notation $\mmean{\hat O}_\Psi=\bra{\psi}\bra{\psi}\hat O\ket{\psi}\ket{\psi}$ to express the mean value for two identical replicas of state $|\psi\rangle$. The second-order moment of $\hat{O}$ gives
\begin{equation}
\mmean{\hat O^2}_\Psi  = \mean{\psi | \hat x^2 | \psi} \, \mean{\psi | \hat p^2 | \psi}.
\end{equation}
In the special case where the distributions of $x$ and $p$ are centered on zero, $\mmean{ \hat O^2}$ thus gives access to the product of variances $\Delta x^2\, \Delta p^2$ in state  $|\psi\rangle$, which is not accessible with a single instance of the state. We may easily verify that the observable $\hat O$ is invariant under a squeezing of the $x$ quadrature with parameter $r$, that is, under the symplectic transformation
\begin{eqnarray}
\hat x \to \hat x^{(r)} = e^{-r}\, \hat x,  \qquad
\hat p \to \hat p^{(r)} = e^{r} \, \hat p .
\end{eqnarray}
Indeed, 
\begin{equation}
\hat O^{(r)} = \hat x^{(r)}_1 \otimes \hat p^{(r)}_2 =  \hat x_1 \otimes \hat p_2 = \hat O  \, ,
\end{equation}
so that measuring $\hat O$ on a state  $\ket{\Psi}$ is insensitive to applying a prior squeezing operation along the $x$ (or $p$) quadrature on state $\ket{\psi}$. However, this property does not extend to rotated states since $\hat O$ is not rotation-invariant.

To fix this problem,  we may use instead of $\hat O$ the uncertainty observable defined as the 2-copy operator 
\begin{equation}
\hat L_z = {1 \over 2}\big(  \hat x_1 \otimes \hat p_2 - \hat p_1 \otimes \hat x_2 \big)
\label{eq-def-L_z}
\end{equation}
where we use index $z$ to denote that it is the third component (or $z$ projection) of an angular momentum $\mathbf{ \hat  L}$.
This definition can be motivated by taking a rotation-averaged version of the above operator $\hat O$. Indeed, using the symplectic transformation for a rotation of angle $\theta$,
\begin{eqnarray}
\hat x^{(\theta)} =  \hat x \, \cos \theta + \hat p \, \sin \theta    ,   \quad \hat p^{(\theta)} = - \hat x \, \sin \theta   +  \hat p \,  \cos \theta   ,
\end{eqnarray}
we have
\begin{equation}
{1\over 2 \pi} \int_0^{2\pi} \hat x_1^{(\theta)} \otimes \hat p_2^{(\theta)} \, d\theta= \, {1 \over 2} \big( \hat x_1 \otimes \hat p_2 - \hat p_1 \otimes \hat x_2 \big).
\end{equation}
This observable is obviously invariant under rotations as well as squeezing operations, hence it is invariant under the set of all symplectic transformations.

The expectation value of ${\hat  L_z}$ vanishes for all states $|\Psi\rangle$, namely
\begin{equation}
\mmean{\hat  L_z}_\Psi = {1 \over 2} \Big( \mean{\hat x}_\psi \mean{\hat p}_\psi - \mean{\hat p}_\psi \mean{\hat x}_\psi \Big) = 0.
\end{equation}
Its second-order moment gives
\begin{eqnarray}
\mmean{\hat L_z^{\,2}}_\Psi &=& {1 \over 2} \Big( \mean{\hat x^2}_\psi \mean{\hat p^2}_\psi - \mean{\hat x \hat p}_\psi \mean{\hat p \hat x}_\psi \Big) \nonumber \\
&=& {1 \over 2} \Big( \mean{\hat x^2}_\psi \mean{\hat p^2}_\psi - {1\over 4} \mean{ \{\hat x , \hat p \} }_\psi^2  + {1\over 4}\mean{[\hat x,\hat p]}_\psi^2  \Big) \nonumber \\
&=& {1 \over 2} \Big( \mathrm{det} \gamma_c  + {1\over 4}\mean{[\hat x,\hat p]}_\psi^2  \Big)
\label{calcVar}
\end{eqnarray}
where we have used the fact that
\begin{eqnarray}
\mean{\hat x \hat p}&=&\frac{1}{2}\Big(\mean{[\hat x,\hat p]}+\mean{\{\hat x,\hat p\}}\Big)\nonumber\\
\mean{\hat p \hat x}&=&\frac{1}{2}\Big(\mean{-[\hat x,\hat p]}+\mean{\{\hat x,\hat p\}}\Big).
\end{eqnarray}
In the last line of Eq.~(\ref{calcVar}), $\gamma_c$ represent the  covariance matrix  of a state $\ket{\psi}$ centered at the origin in phase space and is defined as
\begin{equation}
\gamma_c
=\begin{pmatrix}
\mean{\hat x^2}&\frac{1}{2}\mean{\{\hat x,\hat p\}}\\
\frac{1}{2}\mean{\{\hat x,\hat p\}}&\mean{\hat p^2}
\end{pmatrix}.
\end{equation}
since $\mean{\hat x}=\mean{\hat p}=0$.
Thus,  the variance   of our 2-copy observable $(\Delta \hat L_z)^2=\mmean{\hat L_z^2}-\mmean{\hat L_z}^2=\mmean{\hat L_z^2}$ is linked to the determinant of the covariance matrix $\gamma_c$, namely
\eq{(\Delta \hat L_z)^2=\frac{1}{2}\left(\det \gamma_c+\frac{1}{4}\mean{[\hat x,\hat p]}^2\right).   \label{eq-link-variance-determinant}}
Since a variance must be non-negative, we get
\eq{\det \gamma_c\geq-\frac{1}{4}\mean{[\hat x,\hat p]}^2. \label{varianceLzetGamma}}

If $x$ and $p$ are classical variables, their commutator  vanishes and the symmetrization in the off-diagonal elements of $\gamma_c$ has no effect, hence Eq.~(\ref{varianceLzetGamma}) simply implies that a classical covariance matrix is positive semi-definite. However, if $\hat x$ and $\hat p$ are canonically-conjugate quantum variables, they do not commute ($[\hat x,\hat p]=i$) and Eq.~(\ref{varianceLzetGamma}) is nothing else but the Schrödinger-Robertson uncertainty relation, $\det\gamma\geq\frac{1}{4}$, where $\gamma$ denotes the usual covariance matrix of a state\footnote{Indeed, the covariance matrix $\gamma$ as defined in Eq. (\ref{eq-true-covariance-matrix}) is invariant under displacements which means that $\det\gamma=\det\gamma_c$ for a state centered on the origin.}. 
From this perspective, the Schrödinger-Robertson uncertainty relation simply expresses the inequality  $\mmean{\hat L_z^{\,2}} \geq 0$, where we first need to center the state before measuring $\hat L_z$.
In some sense, this inequality may be deemed trivial as it expresses the fact that the variance of an operator is non-negative. However, its equivalence with the Schrödinger-Robertson uncertainty relation suggests an alternate formulation of the uncertainty relation in terms of the entropy of $\hat L_z$, as analyzed in Sec. 1.G


\subsection{Physical realization of $\hat L_z$}

Let us give a physical interpretation to the 2-copy uncertainty observable  $\hat L_z$.
Using the mode operators $\hat a_j=(\hat x_j+i\hat p_j)/\sqrt{2} $ for $j=1,2$, we may rewrite it as
\begin{equation}
\hat L_z = {i \over 2} \big(  \hat a_1 \hat a_2^\dagger - \hat a_1^\dagger \hat a_2 \big).
\label{Lzaadag}
\end{equation}
From this definition, it is easy to confirm that the action of $\hat L_z$ gives $0$ on any pure Gaussian state centered on the origin (i.e., any squeezed vacuum state). Let $\ket{s}=S(s)\ket{0}$ denote a squeezed vacuum state, where $\ket{0}$ is the vacuum state and $S(s)=e^{\frac{1}{2}(s^*\hat a^2-s\hat a^{\dag2})} $ is the squeezing operator with parameter $s=re^{i\phi}$. Using $\hat a \ket{0}=0$ $\Leftrightarrow  S(s)\hat aS^\dag(s)\ket{s}=0$ $\Leftrightarrow (\cosh r \,\hat a+e^{i\phi}\sinh r\,\hat a^\dag)\ket{s}=0$, we see that $\ket{s}$ satisfies $\hat a \ket{s} = - e^{i\phi}\tanh r\,\hat a^\dag \ket{s}$.
Therefore,
\begin{eqnarray}
\lefteqn{  \hat L_z\ket{s}\ket{s} = \frac{i}{2}(\hat a_1 \hat a_2^\dag - \hat a_1^\dag \hat a_2)\ket{s}\ket{s}  }  \nonumber\\
&=&\frac{i}{2}\Big((-e^{i\phi}\tanh r \, \hat a_1^\dag ) \hat a_2^\dag - \hat a_1^\dag(-e^{i\phi}\tanh r \, \hat a_2^\dag) \Big) \ket{s}\ket{s}\nonumber\\
&=&0.
\label{LzOnSqu}
\end{eqnarray}

	\begin{figure}[]
		\includegraphics[trim = 4.9cm 19cm 4cm 4.5cm, clip, width=0.85\columnwidth]{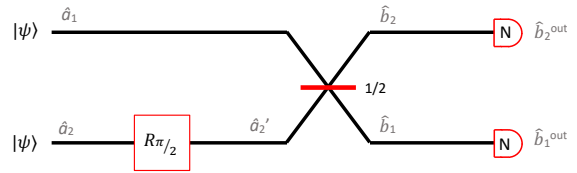}
		\caption{\label{PhysReal} Physical realization of a measurement of the 2-copy uncertainty observable $\hat L_z$. Starting from two identical copies of state $\ket{\psi}$, we apply a $\pi/2$ phase rotation on the second mode and then process the two modes via a 50:50 beam splitter. By measuring the photon number difference of the output state, we access $\hat L_z$. The outcome is zero if and only if $\ket{\psi}$ is a minimum-uncertainty state (Gaussian pure state centered on the origin in phase space).}
	\end{figure}

More interestingly, this formulation of $\hat L_z$ provides us with a nice physical interpretation of the uncertainty observable in terms of a beam-splitter transformation.  As shown in Fig.~\ref{PhysReal}, if we make a $\pi/2$ phase rotation on the second mode,  $\hat a_2\rightarrow \hat a'_2=-i \hat a_2 $, 
followed by a 50:50 beam-splitter transformation of the two modes according to 
\begin{eqnarray}
\hat a_1&\rightarrow&\hat b_1=(\hat a_1 + \hat a'_2)/\sqrt{2}  \nonumber\\
\hat a_2'&\rightarrow&\hat b_2=(\hat a_1 - \hat a'_2)/\sqrt{2}  
\end{eqnarray}
we may reexpress the uncertainty observable as
\begin{equation}
\hat L_z = {1 \over 2} \big(  \hat b_1^\dagger \hat b_1  -  \hat b_2^\dagger \hat b_2   \big)
\label{Lzbbdag}
\end{equation}
where $\hat b_1$ and $\hat b_2$ denote the output mode operators.
Thus, $\hat L_z$ corresponds (up to a factor 1/2) to the difference between the photon numbers at the two output modes of the beam splitter, that is, $\hat L_z=(\hat n_1^{out}-\hat n_2^{out})/2$.


	Remember that a two-mode squeezed vacuum  state can be realized with two single-mode squeezed vacuum states with orthogonal squeezing orientations followed by a 50:50 beam splitter. Thus, if we start with two identical replicas of an arbitrary squeezed vacuum state $\ket{s}\ket{s}$ and rotate one of them by an angle $\pi/2$ before processing both of them through a 50:50 beam splitter, we get precisely a two-mode squeezed vacuum  state. Such a state exhibits perfect photon-number correlations since it is written as $\sum_n c_n \ket{n}\ket{n}$, so measuring the  photon-number difference gives zero with certainty. This is consistent with the fact that our observable $\hat L_z$ gives value $0$ and exhibits no uncertainty (zero variance) when applied to any pure Gaussian state centered on the origin. We have thus found a simple, experimentally relevant method for measuring the uncertainty of a state (or its deviation with respect to a pure Gaussian state\footnote{This method is limited to states centered at the origin in phase space, but we will show in Sect. \ref{Sect-three-copy} how it can be generalized to all states.}).

\subsection{Algebra of angular momenta $(\hat L_x, \hat L_y, \hat L_z)$}

By exploiting the analogy with the algebra of angular momenta, it is possible to define the 2-copy operators $\hat L_x$ and $\hat L_y$, which in turn allows us to define the ladder operators $\hat L_+$ and $\hat L_-$. The definition of  $(\hat L_x, \hat L_y, \hat L_z)$ follows from the Schwinger representation, which yields a connection between an angular momentum and two uncoupled harmonic oscillators (or bosonic modes)  \cite{schwinger}.  In quantum optics, it is also linked to the definition of the Stokes operators in the description of the polarization of light \cite{Collett,bjork,shabbir}. The easiest way to proceed is to note that $\hat L_z$ as defined in Eq.~(\ref{Lzaadag}) can be reexpressed as 
\begin{equation}
\hat L_z=\frac{1}{2}\hat A^\dag\sigma_y \hat A
\label{eq-Lz-Pauli-matrix}
\end{equation}
where $\hat A=\begin{pmatrix}\hat a_1\\\hat a_2\end{pmatrix}$ and  
$\sigma_y=~\begin{psmallmatrix}
0&-i\\i&0
\end{psmallmatrix} $  is the second Pauli matrix.
Similarly, we can define \begin{eqnarray}
\hat L_y=\frac{1}{2}\hat A^\dag\sigma_x
\hat A,\qquad\qquad	\hat L_x=\frac{1}{2}\hat A^\dag\sigma_z \hat A,
\label{eq-Lyx-Pauli-matrix}
\end{eqnarray}
where $\sigma_x=\begin{psmallmatrix}
0&1\\1&0
\end{psmallmatrix}$ and $\sigma_z=\begin{psmallmatrix}
1&0\\0&-1
\end{psmallmatrix} $ are the  other two Pauli matrices. In terms of mode operators or quadrature operators, this gives
\begin{eqnarray}
\hat L_y &=&{1 \over 2}\big(  \hat a_1^\dag  \hat a_2 + \hat a_1  \hat a_2^\dag \big)\nonumber\\
&=&{1 \over 2}\big(  \hat x_1  \hat x_2 + \hat p_1  \hat p_2 \big),\nonumber\\
\hat L_x 
&=&{1 \over 2}\big(  \hat a_1^\dag  \hat a_1  - \hat a_2^\dag  \hat a_2 \big)\nonumber\\
&=&{1 \over 4}\big( ( \hat x_1^2+ \hat p_1^2) - (\hat x_2^2 + \hat p_2^2) \big)\nonumber\\
&=&\frac{1}{2}(\hat n_1-\hat n_2).
\label{lxly}
\end{eqnarray}
Since the Pauli matrices respect the commutation relation $[\sigma_i,\sigma_j]=2i\epsilon_{ijk}\sigma_k$, where $\epsilon_{ijk}$ is the Levi-Civita symbol, it can be verified that our three 2-copy operators respect the commutation relations for angular momenta (see Appendix \ref{AppendixB}) 
\eq{[\hat L_i,\hat L_j]=i\epsilon_{ijk}\hat L_k  . \label{commRela}}
We can then define the ladder operators
\begin{eqnarray}
\hat L_+&=&\hat L_x+i\hat L_y=\frac{1}{2}\big(\hat a_1^\dag+i\hat a_2^\dag\big)\big(\hat a_1+i\hat a_2\big)\nonumber\\
\hat L_-&=&\hat L_x-i\hat L_y=\frac{1}{2}\big(\hat a_1^\dag-i\hat a_2^\dag\big)\big(\hat a_1-i\hat a_2\big).
\label{Ladder}
\end{eqnarray}
as well as the squared angular momentum operator 
\begin{eqnarray}
\hat L^2&=&\hat L_x^2+\hat L_y^2+\hat L_z^2   = \hat L_0 (\hat L_0+1)  
\label{eq-L2}
\end{eqnarray}
where 
\begin{eqnarray}
 \hat L_0 = \frac{1}{2} \big( \hat a_1^\dag\hat a_1 + \hat a_2^\dag\hat a_2 \big) = \frac{1}{2} \left( \hat n_1+\hat n_2 \right)
\label{eq-Casimir}
 \end{eqnarray}
is the Casimir operator.

The definitions of  $\hat L_z$ given in Eqs. (\ref{Lzaadag}) and (\ref{Lzbbdag}) also suggest that all three angular momentum components $(\hat L_x, \hat L_y, \hat L_z)$  can be expressed in alternative ways as a function of the input mode operators $(\hat a_1, \hat a_2)$, output mode operators $(\hat b_1, \hat b_2)$, or even the output mode operators of another circuit $(\hat c_1, \hat c_2)$. This is summarized in Appendix \ref{app_alternative}, together with the corresponding physical realizations of $(\hat L_x, \hat L_y, \hat L_z)$.

\subsection{Eigensystem of $\hat L_z$}
\label{eigensystemSection}

\begin{figure}[h]
	\begin{center}
		\includegraphics[ width=0.6\columnwidth]{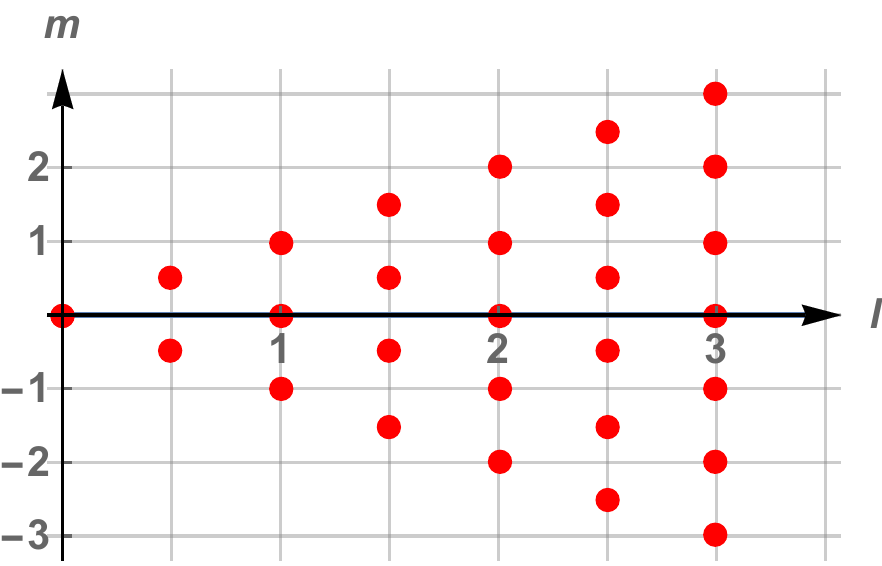}
		\caption{\label{GraphValPropres} Possible eigenvalues $m$ of a state $\kket{l,m}$ with a total photon number equal to $2l$.}
	\end{center}
\end{figure}

In order to calculate the Shannon entropy of the uncertainty observable $\hat L_z$, we need first to determine the eigensystem of this operator.
Defining $l=(n_1+n_2)/2$, we see from Eqs. (\ref{eq-L2}) and (\ref{eq-Casimir})  that the eigenvalues of $\hat L^2$ are given by $l(l+1)$, just as the eigenvalues of the squared modulus of an angular momentum.
Thus, we may label the eigenvectors of $\hat L_z$ and $\hat L^2$ by $\kket{l,m}$, where $l$ represents one half of the total photon number and $m$ is the eigenvalue of $\hat L_z$ (with $|m|\le l$), so that \begin{eqnarray}
\hat L_z \kket{l,m}&=&m\kket{l,m}\nonumber\\
\hat L^2 \kket{l,m}&=&l(l+1)\kket{l,m}\nonumber\\
\hat L_\pm \kket{l,m}&=&\sqrt{l(l+1)-m(m \pm 1)}\kket{l,m \pm 1}  .
\end{eqnarray}
Given the commutation relations (\ref{commRela}),  the possible eigenvalues of $\hat L_z$ for every value of $l$ are $m\in\{-l,l\}$ with integer jumps\footnote{Indeed, $[\hat L_z,\hat L_+]=\hat L_+$ so $\hat L_z\hat L_+\kket{l,m}=(\hat L_+\hat L_z+\hat L_+)\kket{l,m}=(m+1)\hat L_+\kket{l,m}$ where we assumed that $\hat L_z\kket{l,m}=m\kket{l,m}$.} as sketched in Fig.~\ref{GraphValPropres}. 
The eigenvectors of $\hat L_z$ and $\hat L^2$ can be expressed, in general, as linear combinations of the $2$-mode Fock states $\ket{j,k}$,
\begin{equation}
\kket{l,m}=\sum_j\sum_k c_{jk} \ket{j,k}.
\end{equation}

When fixing the value of $l$, the only non-zero $c_{jk}$'s are of course those such that $j+k=2l$.  Let us start with examples for some specific values of $l$.
If we fix $l$=0, the only eigenvector is
\begin{equation}
\kket{0,0}=\ket{0,0}.
\label{kket00}
\end{equation}
If we fix $l=1/2$, we have two eigenvectors with eigenvalues $m=\pm1/2$, namely
\begin{eqnarray}
\kket{1/2,-1/2}&=&\frac{1}{\sqrt{2}}(\ket{0,1}+i\ket{1,0})\nonumber\\
\kket{1/2,1/2}&=&\frac{1}{\sqrt{2}}(\ket{0,1}-i\ket{1,0}).
\end{eqnarray}
If we fix $l=1$, we have three eigenvectors with eigenvalues $m=\{-1,0,1\}$, namely
\begin{eqnarray}
\kket{1,-1}&=&\frac{1}{2}(\ket{2,0}-i\sqrt{2}\ket{1,1}-\ket{0,2})\nonumber\\
\kket{1,0}&=&\frac{1}{\sqrt{2}}(\ket{2,0}+\ket{0,2})\nonumber\\
\kket{1,1}&=&\frac{1}{2}(\ket{2,0}+i\sqrt{2}\ket{1,1}-\ket{0,2}).
\label{kket11}
\end{eqnarray}
For higher values of $l$, it becomes cumbersome to write the general form of the eigenstates but we can, in principle, construct them by applying the ladder operator $\hat L_+$. We start from the eigenstates corresponding to the lowest diagonal  in Fig.~\ref{GraphValPropres}, that is, states $\kket{l,-l}$ whose (unnormalized) form is defined as 
\begin{eqnarray}
\kket{l,-l}&=&\sum_{k=0}^{\lfloor l-1/2\rfloor}\,i^k\sqrt{\binom{2l}{k}}\Big(\ket{k,2l-k}+(-1)^ki^{2l}\ket{2l-k,k}\Big)\nonumber\\
&&+\,\frac{1+(-1)^{2l}}{2}\,i^l\sqrt{\binom{2l}{l}}\ket{l,l}.
\end{eqnarray}
We simply need to apply repeatedly the operator $\hat L_+$ as defined in Eq.~(\ref{Ladder}) in order to find all other eigenstates, since
\begin{equation}
\kket{l,m+1}=\frac{1}{\sqrt{l(l+1)-m(m+1)}}\,\hat L_+\,\kket{l,m}.
\end{equation}
We thus have access to all eigenstates $\kket{l,m}$.

Coming back to the interpretation of $\hat L_z$ as an uncertainty observable, let us discuss the special case of an even total photon number (i.e., when $l$ is an integer). In this case, there is always an eigenstate that admits the eigenvalue $m=0$. Its general (unnormalized) form is
\begin{eqnarray}
\label{mzerostates}
\kket{l,0}&=&\beta\frac{1+(-1)^{l}}{2}\ket{l,l}\\
&&+\sum_{i=0}^{\lfloor l/2-1/2\rfloor}\alpha_i\big(\ket{2i,2l-2i}+\ket{2l-2i,2i}\big)\nonumber
\end{eqnarray}
with
\begin{eqnarray}
\alpha_i&=&\sqrt{\frac{(2l)!!\,(2l-2i-1)!!\,(2i-1)!!}{(2l-2i)!!\,(2l-1)!!\,(2i)!!}}\nonumber\\
\beta&=&\sqrt{\frac{(2l)!!\,(l-1)!!\,(l-1)!!}{(l)!!\,(2l-1)!!\,(l)!!}}
\end{eqnarray}
where $(\cdot)!!$ denotes the double factorial and the index $i$ is an integer.
It means that the states $\kket{l,0}$ are thus written as linear combinations involving only even Fock states of the form $\ket{2j,2k}$. This is connected to the fact that a squeezed vacuum state only involves even Fock states in its expansion. Taking two copies of a squeezed vacuum state $\ket{s}$, namely
\begin{equation}
\ket{s}\otimes\ket{s}=\frac{1}{\cosh r}\sum_{j,k=0}^{\infty}\frac{\sqrt{(2j)!!(2k)!!}}{2^{j+k}j!k!}(\tanh r)^{k+j}\ket{2j,2k} \, ,
\end{equation}
we get again a linear combination of even Fock states $\ket{2j,2k}$. This implies that $\ket{s}\otimes\ket{s}$ can be expressed as a linear combination of eigenstates $\kket{l,0}$ (with $l$ integer). Therefore, applying $\hat L_z$ on $\ket{s}\otimes\ket{s}$ gives zero, which confirms that all squeezed vacuum states $\ket{s}$ are minimum-uncertainty states for the uncertainty observable $\hat L_z$ in accordance with Eq.~(\ref{LzOnSqu}).


Finally, let us mention an interesting symmetry property of the eigenstates $\kket{l,m}$ with respect to the exchange operator $ \hat{P}$, which exchanges the indices of the systems 1 and 2. This operator can be seen as a reflection along the $x_1=x_2$ line and $p_1=p_2$ line in phase space, and it acts on $\hat L_z$, $\hat L_y$ and $\hat L_x$ as follows
\eqarray{\hat P \hat L_z\hat P=-\hat L_z  \qquad
	\hat P \hat L_y\hat P=\hat L_y\qquad
	\hat P \hat L_x\hat P\,=\,-\hat L_x}
where we used  $\hat P^\dag=\hat P$. 
Note also that  $\hat P\hat L_\pm \hat P= - \hat L_\mp$.
Hence, we can evaluate the action of $\hat P$ on the eigenstates of $\hat L_z$. Since $\hat L_z\kket{l,m}=m\kket{l,m}$, we have
\eqarray{-\hat P\hat L_z\hat P\kket{l,m}&=&m\kket{l,m}\nonumber\\
	\Leftrightarrow\qquad\hat L_z\hat P\kket{l,m}&=&-m\hat P\kket{l,m},}
where we used $\hat P^{-1}=\hat P$. Thus, $\hat P\kket{l,m}$ is proportional to the eigenstate of $\hat L_z$ with eigenvalue $-m$, namely
\eq{\hat P\kket{l,m}\propto\kket{l,-m}.\label{SWAP}}
Starting from eigenstate $\kket{l,m}$, we obtain the eigenstate $\kket{l,-m}$ simply by interchanging systems 1 and 2.
From Eq.~(\ref{SWAP}), we also understand that the states $\kket{l,0}$ must be symmetric under the exchange of both systems, as can be checked from Eq.~(\ref{mzerostates}).

\subsection{Entropic uncertainty relation based on $\hat L_z$}

We had seen in Section \ref{sect-1A} that the non-negativity of the variance of our uncertainty observable $\hat L_z$ coincides with the Schrödinger-Robertson uncertainty relation (for states centered at the origin). We will now turn to the Shannon entropy of  $\hat L_z$ and show that it provides a relevant symplectic-invariant  measure of uncertainty. Since we know the eigensystem of  $\hat L_z$ (see Sec. \ref{eigensystemSection}), we can, in principle, compute its Shannon entropy [as defined in Eq. (\ref{shannonentropy})], that is
\eq{H(\hat L_z)_{\rho}=-\sum_{m} p_m\ln p_m  \,  ,}  
where $p_m$ is the probability of measuring eigenvalue $m$ (which goes from $- \infty$ to $\infty$ by steps of 1/2) when having two replicas of state $\rho$, namely
\eq{p_m=\sum_{l=|m|}^\infty \mmean{l,m|\hspace{-2pt} |\rho\otimes\rho|\hspace{-2pt} |l,m}.\label{EqpourCalculpm}}
The sum over $l$ starts at $l=|m|$ since  $-l\leq m\leq l$ and includes only (half-) integer values if $m$ is (half-) integer.

Just like the variance, the Shannon entropy is a non-negative quantity, so it is natural to write
\eq{H(\hat L_z)_{\rho}\geq0  ,  \label{HLz}} 
which is the entropic counterpart of  the Schrödinger-Robertson uncertainty relation $\mmean{\hat L_z^{\,2}} \geq 0$.
It is saturated by all pure Gaussian states (centered at the origin) and is invariant under symplectic transformations (i.e., under any Gaussian unitary except displacements).



Indeed, suppose we apply $U\otimes U$ on an eigenstate $\kket{l,m}$ where $U$ is such a Gaussian unitary. Since $\hat L_z$ is invariant under $U$, i.e., $U^\dag\otimes U^\dag \, \hat L_z \, U\otimes U= \hat L_z$, we have
\begin{eqnarray}
\hat L_z \, U\otimes U\kket{l,m} = m\,U\otimes U\kket{l,m}
\end{eqnarray}
so that $U\otimes U\kket{l,m}$ is an eigenvector of $\hat L_z$ with the same eigenvalue $m$.
Thus, the eigenspace spanned by all states with eigenvalue $m$ is invariant under $U\otimes U$. Hence, the projector associated to the measurement of outcome  $m$
\eq{\mathds{P}_m=\sum_{l=|m|}^\infty\kket{l,m}\bbra{l,m}}
is invariant under $U\otimes U$, and so is the probability of measuring $m$, namely 
$p_m=\tr (\rho\otimes\rho \, \mathds{P}_m)$.
Therefore, the Shannon entropy $H(\hat L_z)_{\rho}$ is invariant under symplectic transformations, as advertised.

\subsection{Special case of Gaussian states}

\begin{figure}[h!]
	\begin{center}
		\includegraphics[ width=0.7\columnwidth]{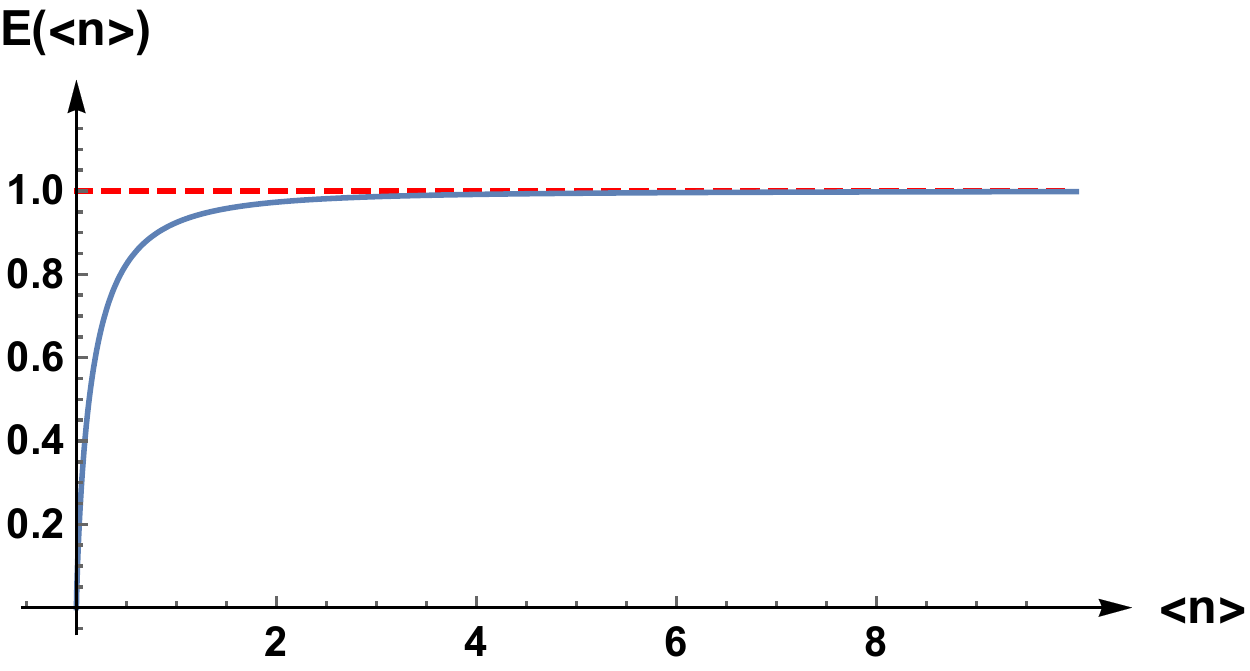}
		\caption{\label{graphEn} Graph of $E(\mean{n})$.}
	\end{center}
\end{figure}

Although it should be easy to measure $\hat L_z$ experimentally (with the circuit in Fig. \ref{PhysReal}) and then compute its Shannon entropy, it does not seem straightforward to calculate $H(\hat L_z)$ analytically for a given state $\ket{\psi}$ because one needs first to express $\kket{\Psi}$ as a linear combination of the eigenstates $\kket{l,m}$.  The calculation of $H(\hat L_z)$ for some simple examples of non-Gaussian states is illustrated in Appendix \ref{exempleNGS}. However, this calculation does not require much effort in the special case of Gaussian states (centered at the origin). Beforehand, remember that, according to Williamson theorem, every Gaussian state can be brought to a thermal state by applying some Gaussian unitary \cite{RMP}. Since $H(\hat L_z)$ is invariant under Gaussian unitaries\footnote{We only consider Gaussian states centered on the origin, which can be brought to a thermal state by applying a symplectic transformation (no displacement is  needed), so the invariance of $H(\hat L_z)$ holds.}, it is enough to compute its  value for a thermal state (it is then the same for any Gaussian state with the same symplectic spectrum). Luckily, it is straightforward to evaluate $H(\hat L_z)$ for a
thermal state
\begin{equation}
\rho_{th}=\sum_{n=0}^{\infty}\frac{\mean{n}^n}{(\mean{n}+1)^{n+1}}\ket{n}\bra{n}
\end{equation}
because when inserting $\rho_{th}\otimes \rho_{th}$ in the circuit of Fig.~\ref{PhysReal}, measuring $\hat L_z$ simply corresponds to measuring the difference between the photon numbers at the two outputs, $\hat d=(\hat n_1^{out}-\hat n_2^{out})/2$. Since a thermal state is invariant under rotation in phase space, the second mode remains in state $\rho_{th}$ after the $\pi/2$ rotation shown in Fig.~\ref{PhysReal}. Moreover, when two copies of a thermal state are inserted in a beam splitter, the output is again the product of the same two thermal states. The random variable $d$ is just the difference of two independent (geometrically distributed) random variables.
The probability of measuring $n_i$ photons on the $i^{th}$ output mode is
\begin{equation}
P(\hat n_i=n_i)=\frac{\mean{n}^{n_i}}{(\mean{n}+1)^{n_i+1}}  \qquad i=1,2
\end{equation}
so the probability of obtaining a certain value for the (half) difference $d$ is 
\begin{equation}
P(\hat d=d)=
\left\{
\begin{array}{lr}
\sum\limits_{n_2=0}^{\infty}P(\hat n_1=n_2+2d)P(\hat n_2=n_2)&d>0\\
\sum\limits_{n_1=0}^{\infty}P(\hat n_1=n_1)P(\hat n_2=n_1-2d)&d<0\\
\sum\limits_{n_1=0}^{\infty}P(\hat n_1=n_1)P(\hat n_2=n_1)&d=0
\end{array}
\right.
\end{equation}
This yields
\begin{equation}
P(\hat d=d)=
\frac{1}{2\mean{n}+1}\left(\frac{\mean{n}}{\mean{n}+1}\right)^{2d},\qquad\forall\,d.
\label{eq-forall_d}
\end{equation}
We can now compute the Shannon entropy of $\hat L_z$ as
\begin{eqnarray}
H(\hat L_z)_{\rho_{th}}&=&-\sum_{d}P(\hat d=d)\ln P(\hat d=d)\nonumber\\
&=&\ln \big( 2\mean{n}+1 \big) + E\big(\mean{n}\big).
\label{eq-entropy-Lz-thermal}
\end{eqnarray}
where
\begin{eqnarray}
E\big(\mean{n}\big) &=& - \frac{2\mean{n}\big( \mean{n}+1\big)}{2\mean{n}+1}\ln\frac{\mean{n}}{\mean{n}+1}
\end{eqnarray}
is a function ranging between 0 and 1, as plotted in Fig.~\ref{graphEn}.
Note that $d$ can be integer or half-integer in Eq. (\ref{eq-forall_d}) and this must be taken into account when summing over $d$ in Eq. (\ref{eq-entropy-Lz-thermal}).
Interestingly, if we compute the Shannon differential entropy $h(x,p)$\footnote{Since a thermal state has a positive Wigner function, its Shannon differential entropy is simply the classical entropy of the joint probability distribution of $(x,p)$ given by the Wigner function.} of a thermal state with Wigner function
\eqarray{W_{\rho_{th}}&=&\frac{1}{2\pi\sqrt{\det \gamma}}e^{-\frac{1}{2}\begin{psmallmatrix}
x&p
		\end{psmallmatrix}^T\gamma^{-1}\begin{psmallmatrix}
		x\\p
	\end{psmallmatrix}}\nonumber\\
&=&\frac{1}{\pi(2\mean{n}+1)}e^{-\frac{1}{2\mean{n}+1}(x^2+p^2)},}
we find
\begin{eqnarray}
h(x,p)_{\rho_{th}}&=&-\int W_{\rho_{th}}(x,p)\ln W_{\rho_{th}}(x,p) \, dx \, dp \nonumber\\
&=&\ln(\pi e)+\ln \big( 2\mean{n}+1\big)
\label{eq-h(x,p)-thermal}
\end{eqnarray}
which implies that
\begin{equation}
H(\hat L_z)_{\rho_{th}}=h(x,p)_{\rho_{th}}-\ln(\pi e)+E(\mean{n}).
\label{eq-interesting-equation}
\end{equation}
This expression is interesting as it combines the Shannon entropy of our discrete uncertainty observable $\hat L_z$ to the Shannon differential entropy of two continuous variables $x$ and $p$. The first term in the r.h.s. of Eq. (\ref{eq-h(x,p)-thermal}) is the Shannon differential entropy of the Wigner function for the vacuum state $h(x,p)_{\rho_{vac}}=\ln(\pi e)$,
so that Eq. (\ref{eq-interesting-equation}) implies that 
$H(\hat L_z)_{\rho_{th}}$ is close to $h(x,p)_{\rho_{th}} - h(x,p)_{\rho_{vac}}$
within a range of  $0\le E\big(\mean{n}\big) \le 1$. This is a way of understanding Eq. (\ref{HLz}) as an entropic uncertainty relation, measuring the distance from a pure Gaussian state (here, the vacuum state).

To be complete, let us also express the above entropies in terms of the symplectic value $\nu$, so this applies to any Gaussian state $\rho_G$. Using the fact that $\mean{n}=\nu-1/2$ for thermal states, we get
\begin{eqnarray}
H(\hat L_z)_{\rho_{G}}&=&\ln(2\nu)-\frac{4\nu^2-1}{4\nu}\ln\frac{2\nu-1}{2\nu+1}\label{HGS}\\
h(x,p)_{\rho_{G}}&=&\ln(\pi e)+\ln (2\nu).
\end{eqnarray}
Note that $H(\hat L_z)_{\rho_{G}}$ is monotonically increasing in $\nu$. The only thermal state that has $H(\hat L_z)=0$ is the vacuum state (considering states centered on the origin). Equivalently, all pure Gaussian states ($\nu=1/2$) saturate our entropic uncertainty relation Eq. (\ref{HLz}), and the quantity $H(\hat L_z)$ can be seen  as a measure of pure non-Gaussianity. Finally, if we only consider Gaussian states, $H(\hat L_z)$ as defined in Eq.~(\ref{HGS}) may also be understood as a measure of mixedness since the purity of a Gaussian state is given by $\mu=\tr{\rho_G^2}=1/2\nu$.

\section{Three-copy uncertainty observable}
\label{Sect-three-copy}

\subsection{Definition of $\hat M$}

The two-copy operator $\hat L_z$ expresses the uncertainty solely for states centered at the origin. To overcome this limitation, we define a 3-copy uncertainty observable, denoted as $\hat M$ in the following. The intuition comes from Ref. \cite{brun}, where it is shown that any $n$th-degree polynomial function of the elements of a single-copy density matrix $\rho$ can be computed as the expectation value of some well-chosen $n$-copy observable acting on $\rho^{\otimes n}$.

We define the covariance matrix $\gamma$ for any state, not necessarily centered on $0$, as
\begin{equation}
\gamma=\begin{pmatrix}
\mean{x^2}-\mean{x}^2&\frac{1}{2}\mean{\{x,p\}}-\mean{x}\mean{p}\\
\frac{1}{2}\mean{\{x,p\}}-\mean{x}\mean{p}&\mean{p^2}-\mean{p}^2
\end{pmatrix}.
\label{eq-true-covariance-matrix}
\end{equation}
This definition is valid for both classical or quantum variables.
If we compute its determinant, we then have
\begin{eqnarray}
\det\gamma
&=&\mean{x^2}\mean{p^2}-\mean{x^2}\mean{p}\mean{p}-\mean{p^2}\mean{x}\mean{x}\nonumber\\
&&-\frac{1}{4}\mean{\{x,p\}}^2+\mean{\{x,p\}}\mean{x}\mean{p}.  	
\end{eqnarray}
From Ref. \cite{brun}, we thus know that this expression must, in principle, be writable as the expectation value of some 4-copy observable. Here, we will show that a 3-copy observable $\hat M$ is actually sufficient if we consider its variance (rather than its expectation value) and follow a similar procedure as for the 2-copy observable $\hat L_z$. As we had seen, the latter is the $z$-component of an angular momentum in the Schwinger representation, but the other two components $\hat L_x$ and $\hat L_y$ are not linked to uncertainty. In contrast, here, we treat the three components $\hat M_i$ of an angular momentum on an equal footing and define\footnote{To be consistent with the definition of the 2-copy observable, we nevertheless introduce a one half factor. This ensures that $\hat M_z=\hat L_z$.} 
\begin{eqnarray}
\hat M_x&=&\frac{1}{2}(	\hat x_2	\hat p_3-	\hat p_2	\hat x_3)\nonumber\\
\hat 	M_y&=&\frac{1}{2}(	\hat x_3	\hat p_1-	\hat p_3	\hat x_1)\nonumber\\
\hat 	M_z&=&\frac{1}{2}(	\hat x_1	\hat p_2-	\hat p_1	\hat x_2).
\label{defLstar}
\end{eqnarray}
The 3-copy uncertainty observable reads
\begin{equation}
\hat M=\frac{1}{\sqrt{3}}(\hat M_x+\hat M_y+\hat M_z)
\end{equation}
and can be viewed as the projection of the angular momentum $\mathbf{\hat M}$ onto a line halfway between the $x$-, $y$-, and $z$-axes.
Since the 2-copy observable $\hat L_z$ is invariant under symplectic transformations (rotations and squeezing), so are all the $\hat M_i$ observables  since they have the same form as $\hat L_z$ acting on two out of the three copies. Hence, the 3-copy observable $\hat M$ is also invariant under symplectic transformations. Furthermore, $\hat M$ is this time also invariant under displacements. Indeed, since we consider three copies of a same state, the displacement is the same in each of the three modes. In other words, the displacement in position $x$ (or momentum $p$) is always applied in the direction  $(\frac{1}{\sqrt{3}},\frac{1}{\sqrt{3}},\frac{1}{\sqrt{3}})$, which is exactly the direction of the angular momentum component $\hat M$. Since the projection of an angular momentum along a direction is invariant under a position shift (or a momentum kick) in that direction, $\hat M$ is invariant under displacements, so we have relaxed the need to restrict to states centered at the origin.

Interestingly,  the variance of $\hat M$ can be related to the determinant of the covariance matrix $\gamma$ exactly as we had done for $\hat L_z$ in Section \ref{sect-1A}.  First, remark that $\mmmean{\hat M}_\Psi =0$, where $\mmmean{\hat M}_\Psi $ stands for the expectation value on three copies of state $\ket{\psi}$. Indeed
\eqarray{\mmmean{\hat M_x}_\Psi &=&\frac{1}{2}\bra{\psi}\bra{\psi}\bra{\psi}\hat M_x\ket{\psi}\ket{\psi}\ket{\psi}\nonumber\\
	&=&\frac{1}{2}(\mean{x}\mean{p}-\mean{p}\mean{x})=0}
and similarly for $\mmmean{\hat M_y}_\Psi $ and $\mmmean{\hat M_z}_\Psi $.
The variance of $\hat M$ is thus equal to its second-order moment, which is computed in Appendix \ref{AppendixC}. We obtain 
\begin{eqnarray}
(\Delta\hat M)^2&=&\mmmean{\hat M^2}\nonumber\\
&=&\frac{1}{3}\mmmean{(M_y+M_x+M_z)^2}\nonumber\\
&=&\frac{1}{2}\mean{x^2}\mean{p^2}-\frac{1}{2}\mean{x^2}\mean{p}\mean{p}-\frac{1}{2}\mean{p^2}\mean{x}\mean{x}\nonumber\\
&&+\frac{1}{2}\mean{\{x,p\}}\mean{x}\mean{p}-\frac{1}{8}\mean{\{x,p\}}^2 +\frac{1}{8}\mean{[x,p]}^2
\nonumber\\
&=&\frac{1}{2}\left(\det\gamma+\frac{1}{4}\mean{[x,p]}^2\right).
\end{eqnarray}
so that the variance of $\hat M$ is related to the determinant of the covariance matrix, in analogy with Eq. (\ref{eq-link-variance-determinant}).
Once again, since a variance is non-negative, we deduce that
\eq{\det \gamma\geq -\frac{1}{4}\mean{[x,p]}^2. \label{gammaComm}}
If $x$ and $p$ are classical, they commute and Eq.~(\ref{gammaComm}) expresses that a covariance matrix is always positive semi-definite. In contrast, if $x$ and $p$ are canonically-conjugate quantum variables, they do not commute ($[x,p]=i$) and Eq.~(\ref{gammaComm}) implies $\det\gamma\geq1/4$, which is the Schrödinger-Robertson relation. This suggests that the 3-copy operator $\hat M$ is a good uncertainty observable, which is invariant under all Gaussian unitaries (including displacements this time). It gives zero with certainty for all Gaussian pure states (regardless of the mean values of $x$ and $p$). We define an entropic uncertainty relation based on the Shannon entropy of this observable
\eq{H(\hat M)_\rho\geq 0.}
As before, to compute the Shannon entropy of $\hat M$, we need to know its eigenvectors and evaluate the associated measurement probabilities. Since $\hat M=(\hat M_x+\hat M_y+\hat M_z)/\sqrt{3}$ is the component of an angular momentum in the direction $(\frac{1}{\sqrt{3}},\frac{1}{\sqrt{3}},\frac{1}{\sqrt{3}})$, its eigenspectrum is well known. More precisely, the eigenvalues of $\hat M_x^2 + \hat M_y^2 + \hat M_z^2$  and $\hat M$ are given, respectively, by
\eqarray{l^*&=&0\qquad\qquad\,\,\, m=0\nonumber\\
	l^*&=&1/2\qquad \qquad m=\left\{-\frac{1}{2},0,\frac{1}{2}\right\}\nonumber\\
	l^*&=&1 \qquad \qquad\,\,\, m=\left\{-1,-\frac{1}{2},0,\frac{1}{2},1\right\}	\nonumber\\
	&\text{etc.}&\qquad}
We do not denote the squared angular momentum operator $\hat M_x^2 + \hat M_y^2 + \hat M_z^2$ simply as $\hat M^2$ here in order to avoid confusion with the square of our uncertainty observable $\hat M$ (which is a component on the angular momentum in a specific direction). Comparing to a genuine angular momentum, the eigenvalues are all divided by two because of the definition of the $\hat M_i$ [see Eq. (\ref{defLstar})]. Moreover, the step between two subsequent eigenvalue is $1/2$ instead of 1 because the commutation relations are $[\hat M_i,\hat M_j]=~\frac{i}{2}\epsilon_{ijk}\hat M_k$ (while it is $[\hat L_i,\hat L_j]=i\epsilon_{ijk}\hat L_k$ for a genuine angular momentum). The eigenfunctions of $\hat M$ are simply the spherical harmonics in the quadrature variables ($x_1,x_2,x_3$), but this form is not very convenient since they must be written in a rotated basis. Computing the probabilities of measuring the eigenvalues of $\hat M$ through the spherical harmonics does not seem to be an easy task, so we find it more suitable to use the physical realization of $\hat M$, see Section \ref{sect-phys-impl-M}.

\subsection{Alternative definitions}
Using the relations between the $x,p$ quadratures and the mode operators, we can express the three angular momentum components as 
\begin{eqnarray}
\hat M_x&=&\frac{i}{2}(\hat a_2\hat a_3^\dag-\hat a_2^\dag \hat a_3)\nonumber\\
\hat M_y&=&\frac{i}{2}(\hat a_3\hat a_1^\dag-\hat a_3^\dag \hat a_1)\nonumber\\
\hat M_z&=&\frac{i}{2}(\hat a_1\hat a_2^\dag-\hat a_1^\dag \hat a_2).
\label{eq-Mx-My-Mz}
\end{eqnarray}
This also allows us to express the squared angular momentum operator as
\eqarray{& \hat M_x^2 + \hat M_y^2 + \hat M_z^2 = \frac{1}{4}\Big((\hat n_1+\hat n_2+\hat n_3)(\hat n_1+\hat n_2+\hat n_3+1)\nonumber\\
	& \left.-(\hat a_1^{\dag2}+\hat a_2^{\dag2}+\hat a_3^{\dag2})(\hat a_1^2+\hat a_2^2+\hat a_3^2)\right)}
where $\hat n_i=\hat a_i^\dag \hat a_i$.
It is symmetric in the modes, but does not have the usual $l(l+1)$ form as we had found for $\hat L^2$ in  Eq. (\ref{eq-L2}). 
Note also that the three components $\hat M_x$, 
$\hat M_y$, and $\hat M_z$ can be written in terms of Gell-Mann matrices, which generalize the Pauli matrices in $3\times3$ dimensions. This makes the counterpart to Eqs. (\ref{eq-Lz-Pauli-matrix}) and (\ref{eq-Lyx-Pauli-matrix}),  see Appendix \ref{app-Gell-Mann}.

\subsection{Physical realization of $\hat M$}
\label{sect-phys-impl-M}

\begin{figure}[tp]
	\begin{center}
		\includegraphics[trim = 4.3cm 16cm 3.8cm 4.2cm, clip, width=0.85\columnwidth]{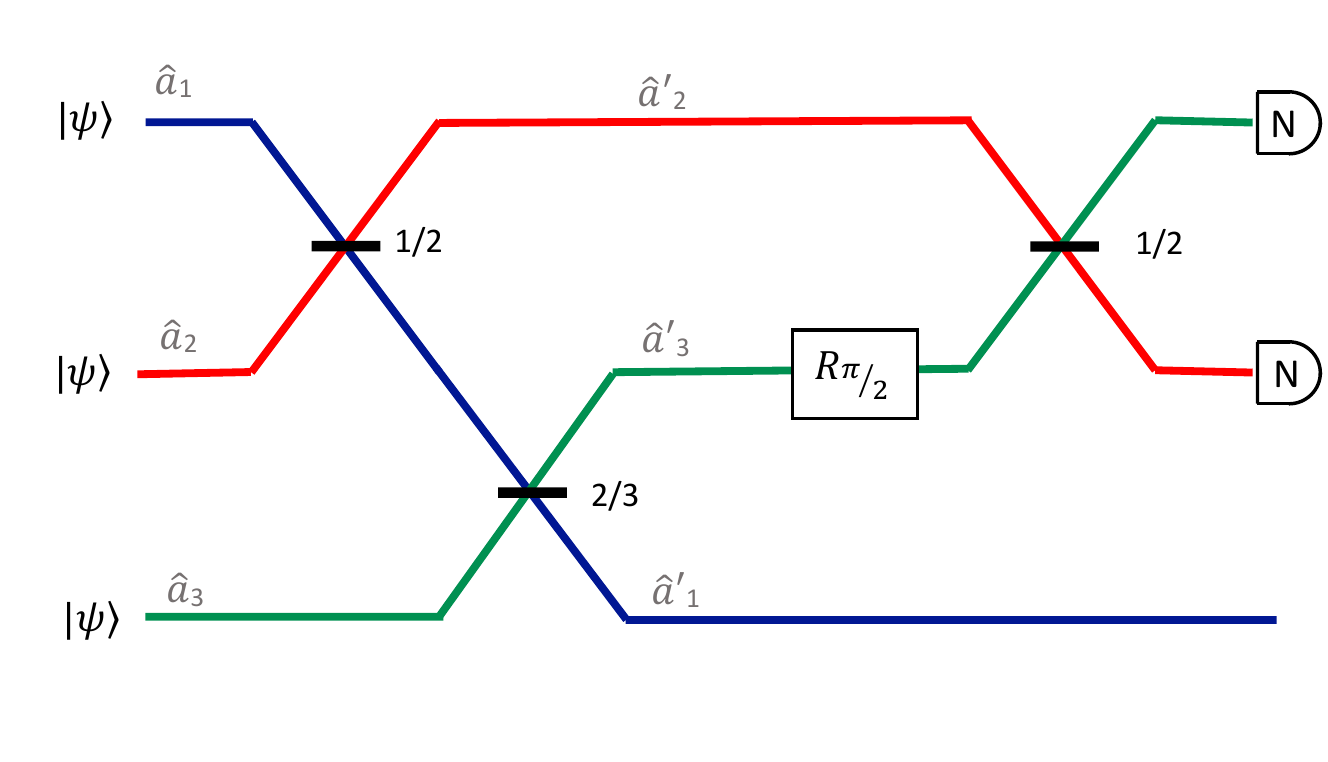}
		\caption{\label{PhysReal3}   Physical realization of a measurement of the 3-copy uncertainty observable $\hat M$. Starting from three identical copies of state $\ket{\psi}$, we first apply two beam splitters (on modes 1 and 2 with transmittance $1/2$, and then on modes 1 and 3 with transmittance $2/3$). This effects a rotation in phase space such that $\hat M$ is rotated towards $\hat M_x$, which is measured by the second part of the circuit consisting of a $\pi/2$ rotation and a 50:50 beam splitter. By measuring the photon number difference of modes 2 and 3, we thus access $\hat M$. The outcome is zero if and only if $\ket{\psi}$ is a minimum-uncertainty state (Gaussian pure state regardless of its position in phase space).}
	\end{center}
\end{figure}

We show in Fig.~\ref{PhysReal3} an optical circuit that allows us to measure the 3-copy uncertainty observable $\hat M$. It is similar to the circuit for the 2-copy observable $\hat L_z$ in the sense that, in the last stage of the circuit, we apply a $\pi/2$ rotation followed by a 50:50 beam splitter and then compute the difference between the output photon numbers. If the circuit was limited to this last stage, the photon number difference on modes 2 and 3 would yield $\hat M_x$, in accordance with the first equation in Eq. (\ref{eq-Mx-My-Mz}) which is analogous to Eq. (\ref{Lzaadag}). However, this transformation is preceded by two beam splitters of transmittance $1/2$ (on modes 1 and 2) and $2/3$ (on modes 1 and 3). The effect of these beam splitters is to make the appropriate rotation in phase space so that the direction  $(\frac{1}{\sqrt{3}},\frac{1}{\sqrt{3}},\frac{1}{\sqrt{3}})$ is turned to $(1,0,0)$, that is, the $x$-direction. Indeed, after applying the two beam splitters, the mode operators are given by  
\eqarray{\hat a_1'&=&\frac{1}{\sqrt{3}}(\hat a_1+\hat a_2+\hat a_3)\nonumber\\
	\hat a_2'&=&\frac{1}{\sqrt{2}}(\hat a_1-\hat a_2)\nonumber\\
	\hat a_3'&=&\frac{1}{\sqrt{6}}(\hat a_1+\hat a_2-2\hat a_3).}
In particular, the first mode operator becomes the sum of the three input mode operators. It  means that measuring the $x$-component angular momentum $\hat M_x$ after this rotation (i.e., on modes $\hat a'_1$, $\hat a'_2$ and $\hat a'_3$) yields the value of  $(\hat M_x+\hat M_y+\hat M_z)/\sqrt{3}$ before the rotation, which is precisely the desired uncertainty observable $\hat M$. Therefore, keeping in mind the analogy with the 2-copy observable $\hat L_z$, we can access $\hat M$ simply by applying a $\pi/2$ rotation followed by a 50:50 beam splitter on modes 2 and 3. The output photon number difference yields
	\eq{\hat M=\frac{1}{2}(\hat n_2^{out}-\hat n_3^{out})=\frac{1}{2}(\hat a'_2\hat {a'}_3^{\dag}-\hat {a'}_{2}^{\dag}\hat a'_3).}

Interestingly, the invariance of $\hat M$ under displacements is easy to understand from the circuit of Fig.~\ref{PhysReal3}. Let us insert a displacement $D(\alpha)$ on each input state of the circuit. After the first two beam splitters, the displacement on the three modes becomes
\eq{D(\alpha)^{\otimes3} \rightarrow D(\sqrt{3}\alpha)D(0)D(0).}
Hence, regardless of the value of $\alpha$, the displacement is zero on modes 2 and 3 just at the point where we apply the $\pi/2$ rotation and the last beam splitter. Therefore, the result of the measurement of the photon-number difference between modes 2 and 3 at the end of the circuit --- which gives $\hat M$ --- does not  depend on the displacement.

Note that we still have a degree of freedom in the state obtained after applying the two first beam splitters in Fig.~\ref{PhysReal3}. Indeed, we can easily verify that applying any real rotation in phase space between modes 2 and 3 (i.e., inserting a beam splitter coupling these modes just before the second part of the circuit) does not affect $M_x$, hence it does not change the measured value of $\hat M$.\footnote{This is related to the fact that $\hat M_x$ is invariant under a real rotation between systems 2 and 3. Indeed, if we define
\eqarray{\hat x_2'=\cos \theta \hat x_2+\sin \theta \hat x_3\qquad
	\hat x_3'=-\sin \theta \hat x_2+\cos \theta\hat x_3}
and similarly for the $p$ quadratures, we can easily show that 
\eq{\hat M_x'=\frac{1}{2}(\hat x_2'\hat p_3'-\hat p_2'\hat x_3')=\frac{1}{2}(\hat x_2\hat p_3-\hat p_2\hat x_3)=\hat M_x.}
}


\subsection{Entropic uncertainty relation based on $\hat M$}

It is easy to verify that our 3-copy uncertainty observable vanishes on any pure Gaussian state (i.e., squeezed coherent state). If we insert three copies of a squeezed coherent state in the optical circuit of Fig.~\ref{PhysReal3}, we obtain after the first two beam splitters the same three squeezed coherent states (albeit with changed mean values, as explained earlier)\footnote{If the product of two identical Gaussian states impinge on a beam splitter, we obtain at the output a product of two Gaussian states with the same covariance matrix (only the mean values are changed).}.  It means that, similarly to the 2-copy case, we get a zero photon-number difference with probability one at the output of the circuit. Consequently, the entropy of $\hat M$ is equal to zero for any pure Gaussian state. Our entropic uncertainty relation $H(\hat M)\geq0$ thus admits the exact same set of minimum uncertainty states as the Schrödinger-Robertson uncertainty relation.

Furthermore, it appears that the entropic uncertainty relation based on $\hat M$ coincides with the one based on $\hat L_z$ in the special case of Gaussian states centered at the origin.
Indeed, if we plug three copies of an arbitrary Gaussian state, pure or mixed, at the input of the circuit of Fig.~\ref{PhysReal3}, we again get the same three Gaussian states after the first two beam splitters (albeit with changed mean values). In particular, we find two copies of the input Gaussian state on modes 2 and 3 (albeit centered at the origin). Since the rest of the circuit is the same as the 2-copy circuit of Fig.~\ref{PhysReal}, all conclusions we had drawn for $\hat L_z$ hold for $\hat M$ too. In particular, the entropy of a Gaussian state will be the same, namely
\eq{H(\hat M)_{\rho^G}=H(\hat L_z)_{\rho^G}}
where $H(\hat L_z)_{\rho^G}$ is defined in Eq. (\ref{HGS}).

In the case of non-Gaussian states centered at the origin, however, we expect the entropy $H(\hat M)$ to deviate from $H(\hat L_z)$, so that it seems relevant to define a distinct entropic uncertainty relation $H(\hat M)\geq0$. For example, if we insert three copies of Fock state $\ket{1}$ in the circuit of Fig.~\ref{PhysReal3}, the state of modes 2 and 3 differs from $\ket{1}^{\otimes2}$ after the first two beam splitters, so the second part of the circuit acts differently. Hence, the entropy of the 3-copy observable $H(\hat M)_{\ket{1}}$ differs from that of the 2-copy observable $H(\hat L_z)_{\ket{1}}$ (as computed in Appendix \ref{exempleNGS}).

\section{Conclusion}

We have paved the way towards the construction of entropic uncertainty relations for continuous-variable bosonic states that are invariant under Gaussian
unitary transformations (rotation, squeezing, and displacement in phase space). This was achieved by defining the notion of multi-copy uncertainty observable (especially a 2-copy observable $\hat L_z$ and a 3-copy observable $\hat M$) with ingrained invariance, building on the Schwinger representation of angular momenta in terms of bosonic operators. Observable $\hat L_z$ acts on two replicas of a continuous-variable state and is invariant under rotation and squeezing (so that it is relevant for states centered at the origin only), while $\hat M$ acts on three replicas and exhibits extra invariance under displacement (so that it is relevant for any state). Expressing the non-negativity of the variance of both (discrete-spectrum) observables $\hat L_z$ and $\hat M$ leads to the Schrödinger-Robertson uncertainty relation, which supports the fact that these observables capture uncertainty in phase space (or the deviation from pure Gaussianity).
Based on this, we have constructed two entropic uncertainty relations by expressing the fact that the Shannon entropy of  $\hat L_z$ and $\hat M$ must be non-negative for any physical state. Given the intrinsic invariance of $\hat L_z$ and $\hat M$, these entropic uncertainty relations are automatically invariant under Gaussian unitaries and are saturated by all pure Gaussian states (with $\hat L_z$ being restricted to states centered at the origin). In some sense, they can be viewed as the entropic counterpart to the Schrödinger-Robertson uncertainty relation.

Although such a Gaussian invariance is not strictly necessary for a measure of uncertainty to be meaningful, if the purpose is to define a measure of uncertainty in phase space rather than a function merely relating the uncertainties of variables $x$ and $p$, it is natural to require this measure to be invariant under symplectic transformations, which leave the volume in phase space invariant. Remarkably, it is the angular-momentum algebra of the uncertainty observables $\hat L_z$ and $\hat M$ that ensures this invariance in our construction.

Next, we have described optical circuits enabling us to measure observable $\hat L_z$ (respectively $\hat M$) starting from two (respectively three) replicas of the input state. From an experimental perspective, measuring these observables requires the preparation of two (or three) identical replicas of an optical state, followed by a linear-optics circuit combining them in order to achieve a specific joint measurement. Thus, the identical optical states should be generated from a same laser (to share a same phase reference) and interferometric stability should be ensured in the optical circuit up to the final measurement of the photon-number difference. The complexity of such a setup is comparable to that of various current experiments on multiphoton interference effects in multimode circuits (in particular those based on integrated photonic chips, see e.g. \cite{crespi}), so it seems reasonable to access the uncertainty $\hat L_z$ or $\hat M$ of a state in $x$-$p$ space, at least when dealing with the optical analogues of $x$ and $p$.

Regardless of the experimental feasibility of measuring observable $\hat L_z$ or $\hat M$, the sole theoretical definition of these optical circuits proved to be useful in order to derive a closed formula for the Shannon entropy $H(\hat L_z)$ and  $H(\hat M)$ in the special case of Gaussian states (both entropies coincide in that case). However, we have not found a simple method to compute these entropies for non-Gaussian states, which we leave as a topic for further study. Another problem that we leave open is to find an operational meaning to $H(\hat L_z)$ and  $H(\hat M)$, which would help interpreting physically the associated entropic uncertainty relations. It is fascinating that the Shannon entropy of a (discrete-spectrum) angular momentum observable such as $\hat L_z$ or $\hat M$ can be connected to the differential entropy of the Wigner function in (continuous-variable) $x$-$p$ space, at least for Gaussian states.

Furthermore, an interesting question raised by this work is to elucidate the reason why three replicas seem to be necessary to build an uncertainty observable that possesses the desired invariance. Since the left-hand side of the Schrödinger-Robertson relation is quartic in the position-momentum variables, the variance of a two-copy observable might have been sufficient (assuming the observable is linear in the quadrature variables of each copy) and it is unclear why we had to consider the variance of a three-copy observable instead (this could, in principle, give access to 6th-order moments in $x$ and $p$). Conversely, a four-copy observable may also have been considered, where some constraint on its mean (e.g., the observable must be positive semi-definite) instead of its variance would induce an uncertainty relation. More generally, a valuable extension of this work would be to investigate general multi-copy uncertainty observables. 


\medskip
\noindent {\it Acknowlegments}: 
This work was supported by the Fonds de la Recherche Scientifique – FNRS under Projects No. T.0199.13 and T.0224.18. AH and OO also acknowledge financial support from the Fonds de la Recherche Scientifique – FNRS.

\appendix



\section{Calculation of the commutator between $\hat L_x$ and $\hat L_y$}
\label{AppendixB}

Let us show that the 2-copy operators $\hat L_x$, $\hat L_y$, and $\hat L_z$ obey the commutation relations for angular momenta. As an example, we calculate the commutator between $\hat L_x$ and $\hat L_y$ using the properties of Pauli matrices, namely
	\begin{eqnarray}
	[\hat L_x,\hat L_y]&=&\frac{1}{4}[A^\dag \sigma_z A,A^\dag \sigma_xA]\nonumber\\
	&=&\frac{1}{4}A^\dag\Big(\sigma_zAA^\dag\sigma_x-\sigma_xAA^\dag\sigma_z\Big)A.
	\end{eqnarray}
	where $\hat A=\begin{pmatrix}\hat a_1\\\hat a_2\end{pmatrix}$.
	We can easily compute
	\eq{A A^\dag=(\hat L_0+1)\mathds{1}+\hat L_y\sigma_x+\hat L_z\sigma_y+\hat L_x\sigma_z}
	where
	\eq{\hat L_0=\frac{\hat a_1^\dag\hat a_1+\hat a_2^\dag \hat a_2}{2}=\frac{1}{2}A^\dag A}
	so that the commutator  becomes
	\eqarray{[\hat L_x,\hat L_y]&=&\frac{1}{4}A^\dag\Big(\sigma_z((\hat L_0+1)\mathds{1}+\hat L_y\sigma_x+\hat L_z\sigma_y+\hat L_x\sigma_z)\sigma_x\nonumber\\
		&&\,\,-\sigma_x((\hat L_0+1)\mathds{1}+\hat L_y\sigma_x+\hat L_z\sigma_y+\hat L_x\sigma_z)\sigma_z\Big)A\nonumber\\
		&=&\frac{1}{4}A^\dag \big((\hat L_0+1)[\sigma_z,\sigma_x]-2i \hat L_z\big)A\nonumber\\
		&=&\frac{i}{2}A^\dag \big((\hat L_0+1)\sigma_y- \hat L_z\big)A\\
		&=&\frac{i}{2}A^\dag \left( \left(\frac{1}{2}A^\dag A+1\right)\sigma_y- \frac{1}{2}A^\dag\sigma_yA\right)A\nonumber\\
		&=&\frac{i}{2}A^\dag\sigma_yA+\frac{i}{4}\left(A^\dag \left( A^\dag A\right)\sigma_y A-A^\dag \left(A^\dag \sigma_yA \right)A\right)\nonumber\\
		&=&i\hat L_z+\frac{i}{4}\left(A^\dag \left( A^\dag A\right)\sigma_y A-A^\dag \left(A^\dag \sigma_yA \right)A\right).\nonumber
	}
	Now, we just need to show that the last term in this expression is equal to zero. However, the calculation is not straightforward because the matrices do not all have consistent dimensions for multiplications\footnote{The matrix multiplication is associative only if we multiply matrices of dimensions $n\times m$, $m\times p$ and $p\times q$.}. Nevertheless, we can prove that
	\eqarray{A^\dag M\left(A^\dag A\right)NA&=&\sum_{ijk} \hat a^\dag_iM_{ij}\left(\sum_l \hat a^\dag_l \hat a_l\right)N_{jk}\hat a_k\nonumber\\
		&=&\sum_l \hat a^\dag_l\left(\sum_{ijk}\hat a^\dag_iM_{ij}N_{jk}\hat a_k\right)\hat a_l\nonumber\\
		&=&	A^\dag\left(A^\dag MNA\right)A}
	where the objects inside the brackets have the dimension of a scalar and the matrices $M$ and $N$ are composed of scalar numbers, so they commute with the mode operators. If we define $M=\mathds{1}$ and $N=\sigma_y$, we have 
	\eq{A^\dag \left( A^\dag A\right)\sigma_y A-A^\dag \left(A^\dag \sigma_yA \right)A=0}
	which completes the calculation of the commutator 
	\eq{[\hat L_x,\hat L_y]=i\hat L_z.\label{commLxLy}}
The other commutators can be calculated similarly.

\section{Alternative definitions of $(\hat L_x,\hat L_y,\hat L_z)$}
\label{app_alternative}

\begin{figure}[tp]
	\begin{center}
		\includegraphics[trim = 3.7cm 13cm 4cm 4.5cm, clip, width=0.75\columnwidth]{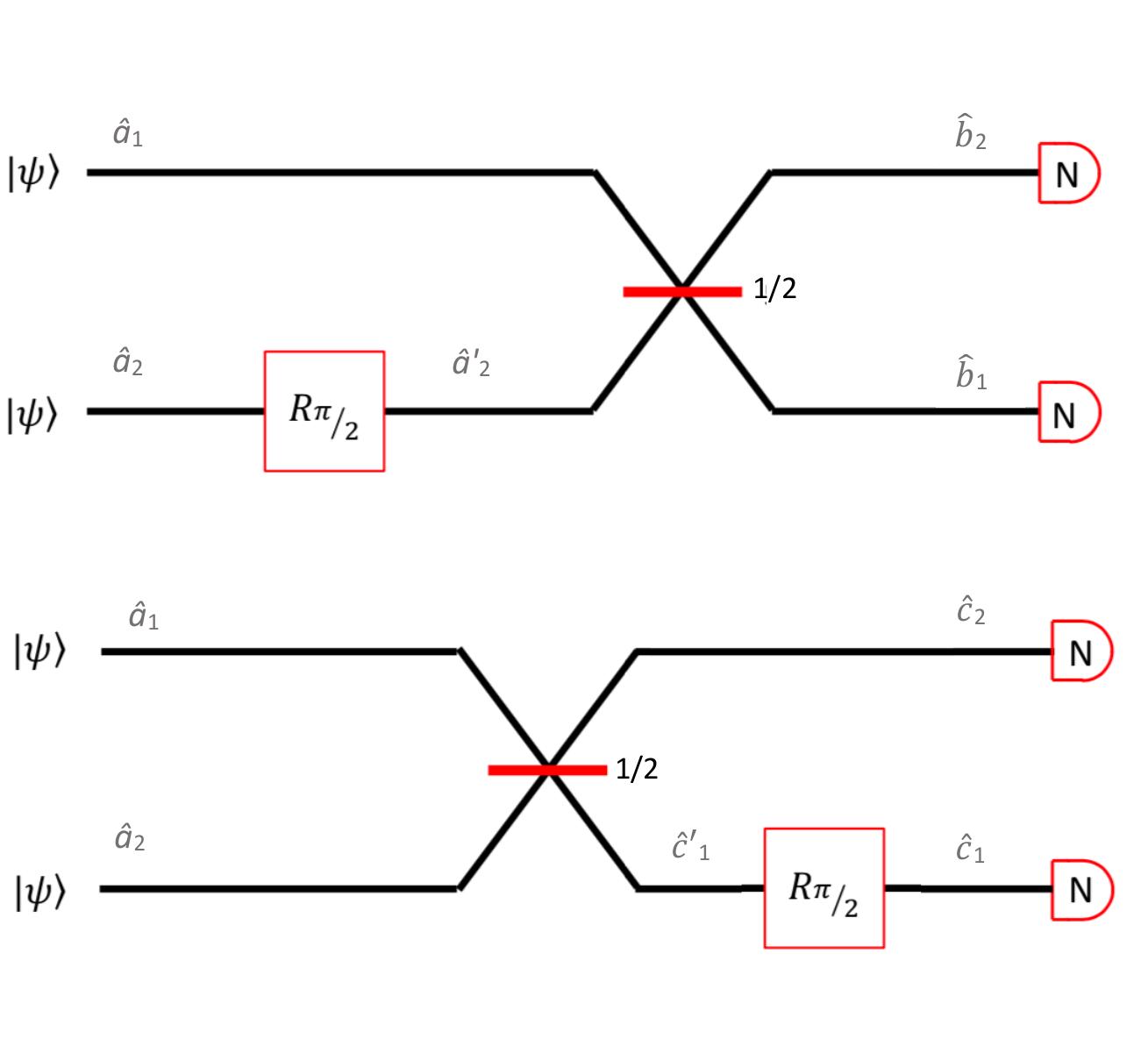}
		\caption{\label{PhysReal2}  Physical realization of a measurement of $\hat L_z$ (upper circuit) and $\hat L_y$ (lower circuit) starting from two identical copies of state $\ket{\psi}$. Input mode operators $\hat a_1,\hat a_2$ are transformed into  output mode operators $\hat b_1,\hat b_2$ in the upper circuit, consisting in a $\pi/2$ phase rotation followed by a 50:50 beam splitter. The photon number difference yields $\hat L_z$.
		Interchanging the $\pi/2$ phase rotation and 50:50 beam splitter leads to  output mode operators $\hat c_1,\hat c_2$ in the lower circuit, so that the photon number difference yields $\hat L_y$. The input photon number difference simply yields $\hat L_x$.  }
	\end{center}
\end{figure}

The angular momentum components $\hat L_x$, $\hat L_y$, and $\hat L_z$  can be expressed in several ways as a function of the input mode operators $(\hat a_1, \hat a_2)$ or output mode operators $(\hat b_1, \hat b_2)$ of the circuit depicted in Figure \ref{PhysReal}, or even the output mode operators $(\hat c_1, \hat c_2)$ of another circuit. This is explained in Fig. \ref{PhysReal2}, where the first circuit is the same as in Figure \ref{PhysReal}. In the second circuit shown in Fig. \ref{PhysReal2}, the $\pi/2$ phase rotation is applied after the 50:50 beam splitter transformation, and the output mode operators are labeled as $\hat c_1$ and $\hat c_2$. The mode operators evolve as
\begin{eqnarray}
\hat a_1&\rightarrow&\hat c_1'=(\hat a_1+\hat a_2)/\sqrt{2},\qquad\quad\hat c_1'\rightarrow\hat c_1=-i\hat c_1'.\nonumber\\
\hat a_2&\rightarrow&\hat c_2=(\hat a_1-\hat a_2)/\sqrt{2},\end{eqnarray}

Let us show that the operators $\hat L_x$, $\hat L_y$ and $\hat L_z$ can 
equivalently be expressed in terms of the $\hat a$, $\hat b$, or $\hat c$ mode operators. In terms of the mode operator $\hat a$, the expressions are given by equations (\ref{Lzaadag}) and (\ref{lxly}). Using the first circuit, we already showed that $\hat L_z$ corresponds to one half the photon-number difference of the output modes, see Eq. (\ref{Lzbbdag}), 
and it is easy to show that 
\begin{equation}
\hat L_x = \frac{1}{2}(\hat b_1\hat b_2^\dag+\hat b_1^\dag \hat b_2)\qquad
\hat L_y = \frac{i}{2}(\hat b_1\hat b_2^\dag-\hat b_1^\dag \hat b_2).
\end{equation}
Based on the second circuit, we can do similar calculations to express $\hat L_x$, $\hat L_y$, and $\hat L_z$ in terms of the mode operators $\hat c$. The  results are summarized in Table \ref{summaryTable}, which also exhibits the expressions of $\hat L_x$, $\hat L_y$ and $\hat L_z$ in terms of the quadrature operators (first row). Moreover, we have
\begin{eqnarray}
\hat L_z&=&\frac12\hat A^\dag \sigma_y\hat A=
\frac12\hat B^\dag\sigma_z\hat B=
\frac12\hat C^\dag\sigma_x\hat C\nonumber\\
\hat L_y&=&\frac12\hat A^\dag \sigma_x\hat A=
\frac12\hat B^\dag\sigma_y\hat B=
\frac12\hat C^\dag\sigma_z\hat C\nonumber\\
\hat L_x&=&\frac12\hat A^\dag \sigma_z\hat A=
\frac12\hat B^\dag\sigma_x\hat B=
\frac12\hat C^\dag\sigma_y\hat C\nonumber\\
\end{eqnarray}
where $\hat A=\begin{pmatrix}\hat a_1\\\hat a_2\end{pmatrix}$,  $\hat B=\begin{pmatrix}\hat b_1\\\hat b_2\end{pmatrix}$ and  $\hat C=\begin{pmatrix}\hat c_1\\\hat c_2\end{pmatrix}$.

%

\renewcommand\arraystretch{2}
\begin{table}[h!]
	\begin{center}
		\begin{tabular}{c||c |c|c}

&$	\hat L_x$&$	\hat L_y$&$	\hat L_z$\\
	\hline
\hline
$\hat x, \hat p$&$\,\frac{(\hat x_1^2+\hat p_1^2)-(\hat x_2^2+\hat p_2^2)}{4}\,$
& $\,\,\frac{1}{2}\left(\hat x_1\hat x_2+\hat p_1\hat p_2\right)\,\,$ & $\frac{1}{2}\left(\hat x_1\hat p_2-\hat p_1\hat x_2)\right)$\\
\hline
$\hat a, \hat a^\dag$ & $\frac{1}{2}(\hat a_1^\dag \hat a_1-\hat a_2^\dag \hat a_2)$
	& $\,\frac{1}{2}(\hat a_1^\dag \hat a_2+\hat a_1\hat a_2^\dag)\,$ &$\frac{i}{2}(\hat a_1\hat a_2^\dag-\hat a_1^\dag\hat a_2)$ \\
\hline
$\hat b, \hat b^\dag$ & $\frac{1}{2}(\hat b_1\hat b_2^\dag+\hat b_1^\dag\hat b_2)$
	&$\,\frac{i}{2}(\hat b_1\hat b_2^\dag-\hat b_1^\dag\hat b_2)\,$ & $\frac{1}{2}(\hat b_1^\dag\hat b_1-\hat b_2^\dag\hat b_2)$\\
\hline
$\hat c, \hat c^\dag$  &$\frac{i}{2}(\hat c_1\hat c_2^\dag-\hat c_1^\dag\hat c_2)$
 &$\,\frac{1}{2}(\hat c_1^\dag\hat c_1-\hat c_2^\dag\hat c_2)$  &$\frac{1}{2}(\hat c_1\hat c_2^\dag+\hat c_1^\dag\hat c_2)$\\
			\hline
		\end{tabular}
		\caption{\label{summaryTable} All  possible definitions of the operators $\hat L_x$, $\hat L_y$, and $\hat L_z$, in terms of the mode operators $(\hat a_1, \hat a_2)$, $(\hat b_1, \hat b_2)$, $(\hat c_1, \hat c_2)$, or quadrature operators $(\hat x,\hat p)$ .}
	\end{center}
\end{table}
\renewcommand\arraystretch{1}

\section{Calculation of  $H(\hat L_z)$ for some examples of non-Gaussian states}
\label{exempleNGS}

We compute here the entropy of the two-copy uncertainty observable $\hat L_z$ for some examples of non Gaussian states.

{\bf Example 1:} Consider the Fock state $\ket{1}$. If we insert two copies of $\ket{1}$ in the optical circuit of Fig. \ref{PhysReal}, we the output state is
\eq{\frac{1}{\sqrt{2}}(\ket{0\,2}-\ket{2\,0}).}
Therefore, the photon-number difference will be $\pm2$, each with probability $1/2$, and the entropy of $\hat L_z$ is given by
\eq{H(\hat L_z)_{\ket{1}}=-\sum_m p_m\ln p_m=-\frac{1}{2}\ln \frac{1}{2}-\frac{1}{2}\ln\frac{1}{2}=\ln 2.}
As expected, this value is greater than 0 since we are dealing with a non-Gaussian state, in agreement with our entropic uncertainty relation Eq. (\ref{HLz}).

{\bf Example 2:} Consider now a mixture of $\ket{0}$ and $\ket{1}$,
\eq{\rho=\alpha\ket{0}\bra{0}+(1-\alpha)\ket{1}\bra{1}.}
Here, we do not use the optical circuit to compute the entropy, but rather Eq. (\ref{EqpourCalculpm}), namely\footnote{Note that there is a slight abuse of notation here since the sum on $l$ takes half-integer steps that is $l=\{|m|,\,|m|+1/2,\,|m|+1,\cdots\}$.}
\eq{p_m=\sum_{l=|m|}^\infty \mmean{l,m|\hspace{-2pt} |\rho\otimes\rho|\hspace{-2pt} |l,m}.}
Since
\eqarray{\rho\otimes\rho&=&\alpha^2\ket{0\,0}\bra{0\,0}+(1-\alpha)^2\ket{1\,1}\bra{1\,1}\\
	&&+\alpha(1-\alpha)\ket{0\,1}\bra{0\,1}+\alpha(1-\alpha)\ket{1\,0}\bra{1\,0}\nonumber}
we only need to consider states $\kket{l,m}$ with $l=\{0,\frac{1}{2},1\}$, which are given in Eqs.~(\ref{kket00})-(\ref{kket11}). Accordingly, the possible values of $m$ are $\{-1,-\frac{1}{2},0,\frac{1}{2},1\}$. We can now compute the different probabilities $p_m$
\eqarray{p_0&=&\sum_{l=0}^{1}\bbra{l,0}\rho\otimes\rho\kket{l,0}\nonumber\\
	&=&\bbra{0,0}\rho\otimes\rho\kket{0,0}+\bbra{1,0}\rho\otimes\rho\kket{1,0}=\alpha^2\nonumber\\
	p_{\pm\frac{1}{2}}&=&\sum_{l=0}^{1}\bbra{l,\pm1/2}\rho\otimes\rho\kket{l,\pm1/2}\nonumber\\
	&=&\bbra{1/2,\pm1/2}\rho\otimes\rho\kket{1/2,\pm1/2}=\alpha(1-\alpha)\nonumber\\
	p_{\pm1}&=&\sum_{l=0}^{1}\bbra{l,\pm1}\rho\otimes\rho\kket{l,\pm1}\nonumber\\
	&=&\bbra{1,\pm1}\rho\otimes\rho\kket{1,\pm1}=\frac{(1-\alpha)^2}{2}}
and the entropy of $\hat L_z$ is equal to 
\eq{H(\hat L_z)_\rho
	=(1-\alpha)^2\ln2-2\alpha\ln \alpha-2(1-\alpha)\ln(1-\alpha)}
which is always greater than 0 except when $\alpha=1$ because then $\rho$ is simply equal to the vacuum state. If $\alpha=0$, we find $H(\hat L_z)_\rho=\ln2$ as expected from Example 1.

Note that the Shannon entropy of this mixture is a concave function of $\alpha$. This suggests that $H(\hat L_z)$ is probably a concave function in general.

\begin{widetext}
\section{Derivation of the second-order moment of $\hat M$}
\label{AppendixC}

To  compute the second-order moment of the 3-copy observable $\hat M$, we first note that
\begin{eqnarray}
(M_x+M_y+M_z)^2&=&M_x^{2}+M_y^{2}+M_z^{2}+M_xM_y+M_yM_x+M_xM_z+M_zM_x+M_yM_z+M_zM_y\nonumber\\
\end{eqnarray}
with 
\begin{eqnarray}
M_x^2+M_y^2+M_z^2
&=&\frac{1}{4}\Big(x_2^2p_1^2 +x_3^2p_1^2+x_1^2p_2^2 +x_3^2p_2^2+ x_1^2p_3^2+ x_2^2p_3^2 \Big)\nonumber\\
&&-\frac{1}{4} \Big(x_1p_1p_2x_2 + x_2p_2p_3x_3 + x_3p_3p_1x_1 +p_1  x_1 x_2p_2+ p_2x_2x_3p_3+ p_3x_3x_1p_1\Big)\nonumber\\
&=&\frac{1}{4}\Big(x_2^2p_1^2 +x_3^2p_1^2+x_1^2p_2^2 +x_3^2p_2^2+ x_1^2p_3^2+ x_2^2p_3^2 \Big)-\frac{1}{8} \Big(\{x_1,p_1\}\{x_2,p_2\}+\{x_2,p_2\}\{x_3,p_3\})\nonumber\\
&&+\{x_3,p_3\}\{x_1,p_1\}\Big)
+\frac{1}{8} \Big([x_1,p_1][x_2,p_2]+[x_2,p_2][x_3,p_3]+[x_3,p_3][x_1,p_1]
\Big)\nonumber\\
\end{eqnarray}
and
\begin{eqnarray}
M_xM_y    +  M_xM_z   +   M_yM_z+M_yM_x+M_zM_x+M_zM_y&=&  -  \frac{1}{2}\left( p_1 p_2 x_3^2  +   p_1 p_3 x_2^2   +  p_2 p_3 x_1^2 \right) -   \frac{1}{2} \left(p_1^2 x_2 x_3 +   p_2^2 x_1 x_3  +  p_3^2 x_1 x_2\right) \nonumber\\
&&   +  \frac{1}{4}  \Big( \{x_1,p_1\}(x_2p_3+p_2x_3)+\{x_2,p_2\}(x_1p_3+p_1x_3)\nonumber\\
&&+\{x_3,p_3\}(x_1p_2+p_1x_2)  \Big) .
\end{eqnarray}
%
Therefore, if we take the mean value of $\hat M^2$ on three copies of the state we obtain
\begin{eqnarray}
\mmmean{\hat M^2}&=&\frac{1}{3}\mmmean{(M_y+M_x+M_z)^2}\nonumber\\
&=&\frac{1}{12}6\mean{x^2}\mean{p^2}-\frac{1}{6}3\mean{x^2}\mean{p}\mean{p}-\frac{1}{6}3\mean{p^2}\mean{x}\mean{x}+\frac{1}{12}6\mean{\{x,p\}}\mean{x}\mean{p}-\frac{1}{24}3\mean{\{x,p\}}^2 +\frac{1}{24}3\mean{[x,p]}^2
\nonumber\\
&=&\frac{1}{2}\left(\det\gamma+\frac{1}{4}\mean{[x,p]}^2\right).
\end{eqnarray}
\end{widetext}

\section{Expression of $\hat M_x$, $\hat M_y$, and $\hat M_z$ in terms of Gell-Mann matrices}
\label{app-Gell-Mann}

Another way of defining the three angular momentum components $\hat M_x$, $\hat M_y$, and $\hat M_z$ relies on the Gell-Mann matrices, which generalize the Pauli matrices in $3\times3$ dimensions. There are eight Gell-Mann matrices, denoted as $\lambda_i$, but we only need three of them, namely
\eqarray{S_x\equiv\lambda_7&=&\begin{pmatrix}0&0&0\\0&0&-i\\0&i&0
	\end{pmatrix},\quad
	S_y\equiv-\lambda_5=\begin{pmatrix}0&0&i\\0&0&0\\-i&0&0
	\end{pmatrix},\nonumber\\
	S_z\equiv\lambda_2&=&\begin{pmatrix}0&-i&0\\i&0&0\\0&0&0
	\end{pmatrix},}
In analogy with Eqs. (\ref{eq-Lz-Pauli-matrix}) and (\ref{eq-Lyx-Pauli-matrix}), we can write the three operators $\hat M_i$ as

\eq{\hat M_x=\frac{1}{2}A^\dag S_xA,\qquad\hat M_y=\frac{1}{2}A^\dag S_yA,\qquad\hat M_z=\frac{1}{2}A^\dag S_zA.\label{FormGellMann}}
where we have defined $\hat A=\begin{pmatrix}
\hat a_1&\hat a_2&\hat a_3
\end{pmatrix}^T$.
From this formulation, we can easily compute the commutation relations between the $\hat M_i$ observables. They almost obey those of an angular momentum, that is
\eq{[\hat M_i,\hat M_j]=\frac{i}{2}\epsilon_{ijk}\hat M_k}
where the $1/2$ factor comes from our definition of the $\hat M_i$ as already mentioned. All the algebraic properties of operators $\hat M_x$, $\hat M_y$, $\hat M_z$ should be describable in a unified form based on  (\ref{FormGellMann}) and the properties of the Gell-Mann matrices.


\end{document}